\documentclass[12pt]{amsart}

\newcommand{\R}{\mathbb R}
\newcommand{\CC}{\mathbb C}
\newcommand{\Imm}{\mathop{\rm Im}\nolimits}
\newcommand{\Sym}{\mathop{\rm Sym}\nolimits}
\newcommand{\Pol}{\mathop{\rm Pol}\nolimits}
\newcommand{\Distrib}{\mathop{\rm Distrib}\nolimits}

\begin{document}
\title[A mathematical interpretation of Feynman integral]
{A mathematical interpretation of the Feynman path integral equivalent to the
formalism of Green functions}
\author{A. V. Stoyanovsky}
\begin{abstract}
We define the notion of distribution on an infinite dimensional space motivated
by the notion of Feynman path integral and by construction of probability measures
for generalized random fields. This notion of distribution turns out to be mathematically
equivalent to the notion of generating functional of Green functions.
\end{abstract}
\email{stoyan@mccme.ru}
\address{Russian State University of Humanities}

\maketitle

\section*{Introduction}

The purpose of this note is to provide a mathematical interpretation of the notion of
Feynman path integral, and to show that it is mathematically
equivalent to the formalism of Green functions.

Our approach to the mathematical interpretation of Feynman path integral is a continuation of
classical physical and mathematical ideas known in this area.
The Feynman path integral expresses a natural philosophy idea
of excitations propagating along all possible (multidimensional in general) trajectories.
The mathematical approach to this idea in the framework of probability theory is based
on the notion of a random process and its multidimensional generalization --- a random field.
This approach leads to the study of measures on infinite
dimensional spaces. Since Feynman integral is not positive real but complex, it seems
reasonable to try to find the notion of distribution on an infinite dimensional space
instead of the notion of measure, so that the Feynman measure be in fact a distribution.
The construction of distributions we looked for should be close to the construction
of probability measures of random fields, i.~e., it should start with a construction for
finite dimensional spaces with the natural compatibility conditions. This way we come
to the conclusion that the space of main functions needed for our theory of distributions is
just the space of polynomials. Such a theory looses the notion of locality, however it
seems rich enough to develop integral calculus necessary for quantum field theory.
This is confirmed by the fact that the formalism of Green functions is naturally included
into such a theory.

Let us describe the contents of the note.

\S1 is devoted to definition of distributions on
infinite dimensional space. In \S2 we define Fourier transform and Green functions
of a distribution,
and describe the space of distributions as dual to the space of continuous polynomial
functionals. In \S3 we consider first examples of distributions and
define real, complex, and pure imaginary Gaussian distributions.

The author is grateful to V.~P.~Maslov for illuminating discussions and to D.~Z.~Kleiman
for constructive critical comments.

\section{Definition of distributions on an infinite dimensional space}

Consider an infinite dimensional real nuclear topological vector space
containing a dense countable subset [1].
For definiteness,
let it be the Schwartz space $S=S(\R^{d+1})$ of real smooth functions rapidly decreasing
at infinity. Let $S'$ be the dual space of tempered distributions. We are going
to give a definition of a (complex valued) distribution
on the space $S'$. This definition will be analogous to
the definition of a measure of cylindrical sets on $S'$ (in another
terminology, a generalized random field on $\R^{d+1}$)
from [1].

Let $\varphi_1$, $\ldots$, $\varphi_n\in S$ be a (possibly linear
dependent) $n$-tuple of functions. They yield
linear functionals on the space $S'$, or a linear map $\pi:S'\to\R^n$.
Conversely, any continuous linear map $\pi:S'\to\R^n$ is obtained in this way
for some $\varphi_1$, $\ldots$, $\varphi_n\in S$.

Denote by $\Pol(\R^n)$
the vector space of complex polynomial functions on $\R^n$. Let
$\Pol(\R^n)'$ be the dual vector space, with the weak topology of the dual space.

For a linear map $\lambda:\R^n\to\R^m$ we have a continuous linear map
\begin{equation}
\lambda_*:\Pol(\R^n)'\to \Pol(\R^m)'.
\end{equation}

By definition, a distribution on the space $S'$ is a set $\mu=(\mu_\pi)$ of elements
$\mu_\pi\in \Pol(\R^n)'$,
one for each linear map $\pi:S'\to\R^n$,
satisfying the following two conditions.

(i) Compatibility with linear maps: for a linear map $\lambda:\R^n\to\R^m$,
we have
\begin{equation}
\mu_{\lambda\circ\pi}=\lambda_*\mu_\pi.
\end{equation}

(ii) Continuousness: $\mu_\pi$ continuously depends on $\varphi_1$, $\ldots$, $\varphi_n\in S$.

We say that a sequence (or a directed set) $\mu^{(r)}$ of distributions
on $S'$ {\it converges} to a distribution $\mu$ if $\mu_\pi^{(r)}\to\mu_\pi$
locally uniformly for all $\pi$.

We symbolically denote the value of the element $\mu_\pi$ on a polynomial
$$
P(s_1,\ldots,s_n)\in\Pol(\R^n)
$$
by
\begin{equation}
\int_{S'} P\left(\int\varphi_1(x)u(x)dx,\ldots,\int\varphi_n(x)u(x)dx\right)D\mu(u).
\end{equation}

Let us denote the topological vector space of distributions on $S'$ by $\Distrib(S')$.

\section{Fourier transform and Green functions}

Fourier transform of a polynomial on $\R^n$ is a linear combination of derivatives
of the Dirac delta function. This yields an isomorphism of vector spaces
\begin{equation}
F:\Pol(\R^n)\to\bigoplus_{k=0}^\infty\Sym_\CC^k({\R^n}'),
\end{equation}
where $\Sym_\CC^k({\R^n}')$ is the $k$-th complex
symmetric power of the dual vector space ${\R^n}'$.

Further, Fourier transform identifies the dual vector space $\Pol(\R^n)'$ with the space
of formal Taylor series on ${\R^n}'$ at zero, which yields an isomorphism
\begin{equation}
F:\Pol(\R^n)'\to\prod_{k=0}^\infty\Sym_\CC^k(\R^n).
\end{equation}
Denote the $k$-th components of these isomorphisms by $F_k$.

For a distribution $\mu$ on $S'$, condition (i) shows that the Taylor series $F\mu_\pi$
are compatible with one another, and applying the Schwartz kernel theorem to the
$k$-th component $F_k$ gives that condition (ii) shows that these Taylor series are
restrictions of a formal Taylor series at zero
$Z=F\mu$ on the space $S$. Taking $k$-th variational derivative, we obtain a linear map
\begin{equation}
F_k=\delta^k|_0F:\Distrib(S')\to\Sym_\CC^k(S'),
\end{equation}
where the $k$-th complex topological symmetric power $\Sym_\CC^k(S')$
consists of complex valued distributions
\begin{equation}
u(x_1,\ldots,x_k)\in S'_\CC(\R^{k(d+1)}), \ \ x_i\in\R^{d+1},\ \ i=1,\ldots,k,
\end{equation}
symmetric with respect to permutations of $x_i$. This yields an isomorphism
\begin{equation}
F=(F_k):\Distrib(S')\simeq\prod_{k=0}^\infty\Sym_\CC^k(S').
\end{equation}
We symbolically denote the value of
$F_k\mu$ on a main complex function
\begin{equation}
g(x_1,\ldots,x_k)\in S_\CC(\R^{k(d+1)}), \ \ x_i\in\R^{d+1},
\end{equation}
by
\begin{equation}
\begin{aligned}{}
&\int_{S'}\!\left.\int g(x_1,\ldots,x_k)
\frac{\delta^k}{\delta\varphi(x_1)\ldots\delta\varphi(x_k)}
dx_1\ldots dx_k
e^{i\int\varphi(x)u(x)dx}D\mu(u)\right|_{\varphi=0}\\
&=i^k\int_{S'}\!\int g(x_1,\ldots,x_k)u(x_1)\ldots u(x_k)dx_1\ldots dx_kD\mu(u).
\end{aligned}
\end{equation}
The distribution
\begin{equation}
F_k\mu=i^k\int_{S'}u(x_1)\ldots u(x_k)D\mu(u)
\end{equation}
is called the {\it $k$-th Green function} of the distribution $\mu$ on $S'$,
and the Taylor series
\begin{equation}
Z(\varphi)=F\mu(\varphi)=\int_{S'}
e^{i\int\varphi(x)u(x)dx}D\mu(u)
\end{equation}
on $S$ is called the
{\it generating functional of the Green functions} of the distribution $\mu$.

Similarly, using the Schwartz theorems on tensor powers of the space $S$, we obtain that
Fourier transform yields an isomorphism
\begin{equation}
F:\Pol(S')\simeq\bigoplus_{k=0}^\infty\Sym_\CC^k(S)
\end{equation}
of the space $\Pol(S')$ of continuous polynomial functionals on the space $S'$
with the natural topology
with the direct sum of complex topological symmetric powers $\Sym_\CC^k(S)$
of the space $S$, consisting of functions (9) symmetric with respest to
permutations of variables $x_i$. Combining (8)
and (13), we obtain an isomorphism of topological vector spaces
\begin{equation}
\Distrib(S')\simeq\Pol(S')'.
\end{equation}

{\bf Remark.} It is obvious how to generalize the constructions of this paper to
fermionic fields: one should replace everywhere symmetric algebras by Grassmann algebras
consisting of skew symmetric functions.

\section{First examples of distributions}

\subsection{A measure of cylindrical sets}
Let $\mu=(\mu_\pi)$ be a measure of cylindrical sets on
$S'$ in the sense of [1] such that for any $\pi:S'\to\R^n$ the corresponding measure $\mu_\pi$
on $\R^n$ is a continuous density rapidly decreasing at infinity supported on
a vector subspace of $\R^n$. Assume that $\mu_\pi$ continuously depends on $\pi$
in the sense of the present paper and in the sense of [1].
For example, $\mu$ can be a Gaussian measure on $S'$ (possibly degenerate, i.~e., supported
on a closed subspace of $S'$). Then
$\mu$ yields a distribution on $S'$. In [1] it is shown that such a measure is always
countably additive.

\subsection{The complex Gaussian distribution} Let $B(\varphi,\psi)$ be a symmetric bilinear
complex continuous functional on the space $S$ with a positive definite imaginary part:
\begin{equation}
\Imm B(\varphi,\varphi)>0.
\end{equation}
Then we can repeat the construction of a Gaussian measure from [1] in this more general case
to obtain the corresponding complex Gaussian distribution, denoted symbolically
by
\begin{equation}
\sqrt{\det\frac{iB^{-1}}{2\pi}}\ e^{-\frac i2B^{-1}(u,u)}Du,
\end{equation}
on $S'$. To make the paper more self-contained, let us
recall this construction.

For a linear map $\pi:S'\to\R^n$ corresponding to $\varphi_1$, $\ldots$, $\varphi_n\in S$,
we have the dual linear map $\pi':{\R^n}'\to S$. It is well known that due to (15), the
restriction of the bilinear form $B$ to the image of this map is non-degenerate. Let us
choose a basis $\psi_1$, $\ldots$, $\psi_k$ in this image, let us identify it
with ${\R^k}'$, so that we have the decomposition
\begin{equation}
\pi=\rho\circ\pi_1:S'\to\R^k\to\R^n,
\end{equation}
the map $\rho:\R^k\to\R^n$ being injective, and the map $\pi_1:S'\to\R^k$ being surjective,
and consider the linear functional $\mu_{\pi_1}$ on $\Pol(\R^k)$ given by
\begin{equation}
P(s_1,\ldots,s_k)\mapsto\sqrt{\det\frac{iB^{-1}}{2\pi}}
\int_{\R^k}P(s)e^{-\frac i2B^{-1}(s,s)}ds_1\ldots ds_k.
\end{equation}
Here $s=(s_1,\ldots,s_k)$, $B^{-1}$ is the dual bilinear form to $B|_{{\R^k}'}$ on the space
$\R^k$, and the square root of the determinant is chosen to be continuous on the space of
matrices with positive definite real part and to satisfy $\sqrt{\det 1}=1$, so that
$\mu_{\pi_1}(1)=1$. Further,
put $\mu_\pi=\rho_*\mu_{\pi_1}$. We leave to the reader the check of conditions (i) and (ii)
of a distribution on $S'$.

\subsection{The pure imaginary Gaussian distribution} Let $B(\varphi,\psi)$ be a real continuous
symmetric bilinear form on $S$. Adding to it a form $i\varepsilon(\varphi,\psi)$, where
$\varepsilon$ is a positive definite form, we get a form $B+i\varepsilon$, and we can
assign the Gaussian distribution (16) on $S'$ to it. Now, as $\varepsilon\to0$, this
Gaussian distribution tends to a distribution on $S'$ called the pure imaginary Gaussian
distribution. Details are left to the reader.


\begin{thebibliography}{9}
\bibitem{1} I. M. Gelfand, N. Ya. Vilenkin, Generalized functions, vol.~4. Some applications
of harmonic analysis. Equipped Hilbert spaces. Fizmatlit, Moscow, 1961 (in Russian).
\end{thebibliography}
\end{document}